\documentclass[12pt,showpacs,amsmath,amssymb]{revtex4}
\usepackage{bm}
\usepackage{dcolumn}
\usepackage{graphicx}
\usepackage{epstopdf}
\usepackage{color}
\usepackage{epsf}
\usepackage{epsfig}

\newlength{\myhspace}
\newcommand{\hs}{\hspace{\myhspace}}
\begin{document}

\title{Cosmic Jets}

\author{C. Chicone}

\affiliation{Department of Mathematics and Department of Physics and Astronomy, University of Missouri, Columbia,
Missouri 65211, USA }

\author{B. Mashhoon}

\affiliation{Department of Physics and Astronomy, University of Missouri,
Columbia, Missouri 65211, USA}

\author{K. Rosquist}

\affiliation{Department of Physics, Stockholm University, 106 91 Stockholm, Sweden\\and\\
International Center for Relativistic Astrophysics Network (ICRANet), Piazza della Repubblica 10, 65122 Pescara, Italy}

\begin{abstract} We discuss time-dependent gravitational fields that  ``accelerate" free
test particles to the speed of light resulting in cosmic double-jet
configurations. It turns out that complete gravitational collapse along a
spatial axis together with corresponding expansion along the other two axes
leads to the accelerated motion of free test particles up and down parallel
to the collapse axis such that a double-jet pattern is asymptotically formed
with respect to the collapsed configuration.
\end{abstract}

\maketitle
\section{introduction}
In recent papers on gravitomagnetic jets~\cite{1}, a Ricci-flat solution of Einstein's gravitational field equations has been employed to show that in a proper spacetime region devoid of matter and singularities, it is possible for free test particles to ``accelerate" under gravity alone so as to approach the speed of light. The spacetime metric is  in this case~\cite{1}
\begin{equation}\label{eq:1}
ds^2=e^t d\Sigma^2+e^{-t} dz^2,
\end{equation}
where $d\Sigma^2$ is a  stationary 3D  metric given by
\begin{equation}\label{eq:2}
d\Sigma^2=\frac{X_r}{X}(-X^2 dt^2+\frac{1}{r^3} dr^2) +\frac{1}{r}(-Xdt +d\phi)^2.
\end{equation}
The spacetime metric has signature $+2$ and we use units such that $c=1$ throughout.
In Eq.~\eqref{eq:2},  $X_r=dX/dr$ and $X$ is an appropriate solution of the differential equation
\begin{equation}\label{eq:fode}
r^2 X^2 X_{rr}+X_r=0.
\end{equation}

In this rotating cylindrically symmetric spacetime, the geodesics for a fixed value of the radial coordinate $ r$ generally propagate up and down along helical paths around the rotation axis $z$ and their speeds eventually approach the speed of light as $t\to\infty$. These gravitomagnetic jets are attractors. The motion of such free test particles is measured with respect to  the congruence of \emph{fundamental} observers in the background spacetime domain; by definition, these hypothetical test observers are all at rest in space. The introduction of such ``comoving"  observers is in keeping with the way the speeds of astrophysical jets are directly measured. 

   A common feature of quasars and active galactic nuclei is the occurrence of relativistic outflows in the form of a double jet propagating up and down, presumably along the rotation axis of a central ``engine.''  Similar phenomena are also observed, but on a much smaller scale, in microquasars, which are certain X-ray binary systems in our galaxy. The speeds of such astrophysical jets can be directly determined from the rate of movement of plasma clumps in the jet relative to certain more permanent features of the ambient medium.

The purpose of this \emph{Letter} is to demonstrate the occurrence of cosmic double-jet configurations in certain simple time-dependent solutions of Einstein's equations that involve simultaneous collapse along a spatial axis and expansion along the other axes. Therefore, we examine the motion of free test particles with respect to the comoving observers of the background dynamic spacetime; that is, measurements are performed by a fundamental family of comoving observers that are  all at rest in space with four-velocity vector 
\begin{equation}\label{eq:4}
U^\mu=(-g_{tt})^{-1/2}\delta ^\mu_{\hs 0}.
\end{equation}
Here only positive square roots are considered throughout. For these special observers, the corresponding tetrad frame is   $\lambda^\mu_{\hs (\alpha)}$, $\lambda^\mu_{\hs (0)} = U^\mu$,        and the four-velocity vector of a free particle as determined by the background comoving observers along its world line is then
\begin{equation}\label{eq:5}
u^{(\alpha)}=u^\mu \lambda _\mu^{\hs (\alpha)}:=\gamma(1,\mathbf{v}).
\end{equation}
As $ t\to\infty$     along the special geodesics in~\cite{1},   $v^2\to 1$   and $\gamma\to \infty$; indeed, this divergence is an observer-independent property as a consequence of local Lorentz invariance, and is the central feature of gravitomagnetic jets.     

 When    $\gamma\to \infty$,  the free test particle has gained --- in an invariant sense --- an enormous amount of energy from the background gravitational field. A pointwise discussion of gravitational energy is problematic within general relativity due to Einstein's local principle of equivalence. Moreover, since the main issues here involve the ultrarelativistic regime, there is no Newtonian analog that could help clarify the situation. In any case, it should be pointed out that for an extremely energetic particle with   $\gamma\to\infty$, the test particle approximation breaks down; therefore, background fields that exhibit this property are in a certain sense unstable. 

   Inspection of how the ``acceleration"  to the speed of light comes about in solution (1)--(3) reveals that the ``collapse" along the rotation axis $z$  and the simultaneous expansion along the other spatial axes are crucial.  To verify this interpretation, we consider the Kasner spacetime  in detail and demonstrate the efficacy of this concept. 
   
\section{Kasner spacetime}
 Let us study the behavior of timelike geodesics in the standard Kasner metric~\cite{9}
\begin{equation}\label{eq:6}
ds^2=-dt^2+t^{2p_1}dx^2+t^{2 p_2} dy^2+t^{2 p_3} dz^2,
\end{equation}\begin{equation}\label{eq:7}
p_1+p_2+p_3=p_1^2+p_2^2+p_3^2=1.
\end{equation}
We assume that    $p_1<p_2<p_3$;                    that is,
\begin{equation}\label{eq:8}
-\frac{1}{3}\le p_1\le 0,\qquad 0\le p_2\le \frac{2}{3},\qquad \frac{2}{3}\le p_3\le 1.
\end{equation}
In this prototypical expanding anisotropic cosmological model, coordinates are admissible for $t  \in (0,\infty)$,              $t = 0$  at the singularity                    and  $(-g)^{1/2}=t$.                 The Ricci-flat Kasner solutions of the Einstein field equations are indispensable in the theoretical studies of the behavior of generic cosmological models on approach to the singularity~\cite{10,11}.  The Kasner models~\cite {12} are algebraically general (Petrov type I) and spatially homogeneous (Bianchi type I).
 
 Imagine a timelike geodesic world line with unit tangent vector  $u^\mu=dx^\mu/d\tau$, where $x^\mu=(t,x,y,z)$.       It follows from the existence of the Killing vectors  $\partial_x$,  $\partial_y$    and $\partial_z$       that the components of the four-velocity vector of a geodesic along these Killing vectors are constants of the motion; that is, 
  \begin{equation}\label{eq:9}
  t^{2 p_1}\frac{dx}{d\tau}=C_1,\qquad   t^{2 p_2}\frac{dy}{d\tau}=C_2,\qquad t^{2 p_3}\frac{dz}{d\tau}=C_3,
\end{equation}
where $C_1$, $C_2$ and $C_3$ are constants. 
From $u^\mu u_\mu = -1$ and the assumption that $t$ increases with proper time $\tau$ along the geodesic, we find 
  \begin{equation}\label{eq:10}
\frac{dt}{d\tau}=(1+C_1^2 t^{-2 p_1}+C_2^2 t^{-2 p_2}+C_3^2 t^{-2 p_3})^{1/2}.
\end{equation}
For     $C_1=C_2=C_3=0$,   we have geodesic ``observers"  that are comoving with  the Kasner background. These fundamental observers (that are spatially at rest) have orthonormal tetrads   $\lambda ^\mu_{\hs (\alpha)} $, where $\lambda ^\mu_{\hs (0)}=U^\mu=\delta^\mu_{\hs 0} $ and
\begin{equation}\label{eq:11}
\begin{split}
\lambda ^\mu_{\hs (1)} = (0,t^{-p_1},0,0),\quad \lambda ^\mu_{\hs (2)} = (0,0,t^{-p_2},0),\\
\lambda ^\mu_{\hs (3)} = (0,0,0,t^{-p_3}). \hspace{.75in}
\end{split}
\end{equation}

\begin{figure}
\includegraphics[scale=0.8,angle=0]{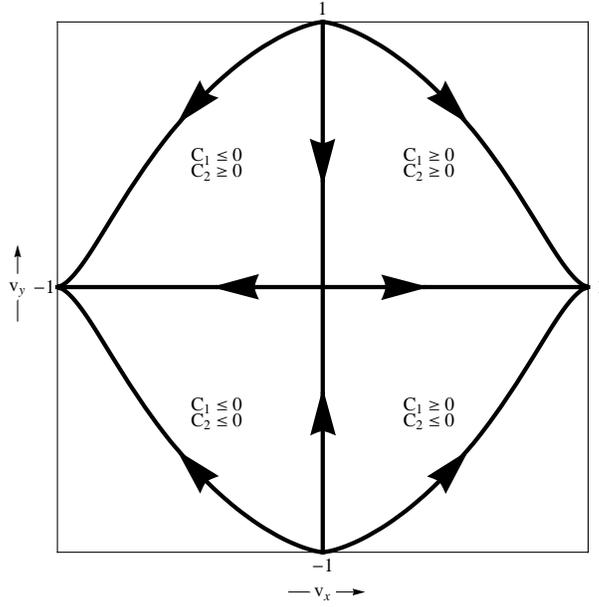}
\caption{If in Eq.~\eqref{eq:12}, we let $C_3=0$ and consider $p_1=-1/3$, $p_2=p_3=2/3$; then, $v_x=C_1 t^{1/3}/\gamma$,  $v_y=C_2 t^{-2/3}/\gamma$ and $v_z=0$, where $\gamma=(1+C_1^2 t^{2/3}+C_2^2 t^{-4/3})^{1/2}$.
These are plotted here for  $C_1=\pm 1$ and $C_2=\pm 1$. The arrows represent the direction of $ t\!: 0\to \infty$ for various combinations of $(C_1,C_2)$  as indicated in each quadrant. We note that    $v_x=\pm 1$  corresponds to a cosmic double-jet configuration that is a \emph{speed-of-light attractor}, while $v_y=\pm 1$ corresponds to a configuration that is a \emph{speed-of-light  repellor}.  \label{fig:1} }
\end{figure}

We are interested in the motion of free test particles relative to the ``ambient medium,'' characterized here by the hypothetical set of test observers with $(C_1  , C_2  , C _3 ) = 0$.  Therefore, we assume that  $(C_1  , C_2  , C _3 ) \ne  0$  for free test particles under consideration here. Then, for the speed of such a particle 
with respect to the fundamental observers,  one finds from Eq.~\eqref{eq:5} that $\gamma=dt/d\tau$, which is given by Eq.~\eqref{eq:10},  and 
 \begin{equation}\label{eq:12}
\gamma v_x=C_1 t^{-p_1},\quad  \gamma v_y=C_2  t^{-p_2},\quad \gamma v_z=C_3  t^{-p_3}.
\end{equation}
It is now straightforward to see that as  $t\to\infty$,  $v_x\to C_1/|C_1|$, $v_y\to 0$ and $v_z\to 0$, which defines a double-jet configuration.
Moreover, for the class of particles with $C_1= 0$, $ \mathbf{v}\to  0$   as $t \to \infty$ . 
Except for this measure-zero set of geodesic particles that are asymptotically at rest with respect to the background comoving observers (``ambient medium"), 
free test particles in Kasner spacetime for $t\to \infty$ move ultrarelativistically, up to the speed of light, up and down parallel to the direction in which space is contracting, while
 parallel to the expanding directions, the corresponding speeds tend to zero. We call the resulting double-jet configuration a \emph{speed-of-light  attractor}.   The dynamical system consisting of timelike geodesics does not have an attractor (or repellor) in the usual sense,  as the speed of light is beyond the reach of a timelike geodesic test particle.  Let us note that some of  the notions illustrated here for $ t\to \infty$  apply equally well for $t\to0$. Indeed, it follows from Eq.~\eqref{eq:12} that for $t\to 0$, 
\begin{align}
\label{eq:13}
v_x\to 0, \qquad v_y^2+v_z^2\to 1.
\end{align}
That is, as $t\to  0$, there is expansion along the $x$  direction and contraction along the other two spatial directions in Eq.~\eqref{eq:6}; therefore, free test particles move with speeds that approach the light speed at early times as well, with their speeds vanishing along the expanding direction.   On the other hand, the $t \to 0$ configuration is a repellor. These aspects of
free particle motion in Kasner spacetime are illustrated in Figure~\ref{fig:1}.

We expect the existence of spacetimes  with  local speed-of-light attractors.  That is, as $t\to \infty$, $\gamma\to \infty$ with respect to the ambient medium is a feature that would be restricted to a congruence of timelike geodesics instead of all timelike geodesics not in the ambient medium as in the Kasner spacetime.

\section{Kasner-like solution with matter}
There are many Kasner-like solutions of general relativity with matter --- see chapter 13 of Ref.~\cite{12}. The two cases of gravitational ``acceleration''  that  we have discussed thus far occur in  vacuum spacetimes; therefore, it is interesting to consider a Kasner-like solution with matter. 

It turns out that Kasner metric~\eqref{eq:6} is consistent with the existence of a comoving perfect fluid source of density $\rho$ and pressure $P$ provided
 \begin{equation}\label{eq:14}
 p_1+p_2+p_3=1, \qquad p_1^2+p_2^2+p_3^2=Q^2\le 1.
\end{equation}
Then,
\begin{align}\label{28}
P=\rho=\frac{1-Q^2}{16 \pi G t^2},
\end{align}
so that for $Q^2  = 1$ the Ricci-flat Kasner spacetime is recovered. In connection with the necessity of the stiff equation of state   $P=\rho$  in this case, see chapter 11 of Ref.~\cite{12} and the references cited therein. 
   
The geodesics in this case behave exactly as before if we make sure that $p_1 < 0$,  $p_2 > 0$ and $p_3 > 0$.  For instance, let $p_2 = p_3$ for simplicity; then, it is possible to  assume that
\begin{equation}\label{eq:15}
p_1=-\frac{1}{3}+2\eta,\quad p_2=p_3=\frac{2}{3}-\eta,
\end{equation}
where $\eta$ is given by
\begin{align} \eta=\frac{1}{3}-\Big( \frac{3 Q^2-1}{18}\Big)^{1/2}.\end{align}
Hence, $p_1 < 0$ and $p_2 = p_3 > 0$ if  $Q^2 > 1/2$;  therefore,  we choose $Q^2$ such that $1 > Q ^2> 1/2$ in Eq.~\eqref{28}. We note that the background perfect fluid in this case plays the role of the fundamental observers. 

To provide a physical interpretation of this result,  imagine particles of the background ambient medium at rest along the $x $ axis. In time, the proper distance between these particles shrinks to zero, while in a similar situation the proper distance between background particles along the $y$ (or $z$) axis expands to infinity; that is, in time, the ambient medium collapses along the $x$ axis to a pancake-like configuration in the $(y, z)$  plane. However, free test particles that are not part of the background medium and initially move with nonzero velocity relative to the background medium get accelerated to the speed of light parallel to the shrinking $x$  axis, forming a jet-like configuration up and down parallel to this axis as they lose their velocities along the other axes. The connection to actual jets would probably involve MHD considerations to confine and collimate the particles where magnetic fields are strong, while in other regions such energetic particles simply escape. 
   
As mentioned before, the features we have discussed persist even in the opposite limit of $t\to 0$; indeed, the Kasner metric approaches a cigar-like configuration as the $x$ axis expands while the $y$ and $z$ axes shrink. Then,   $v_x\to 0$  while $\gamma\to \infty$. The unstable cigar-like configuration thus undergoes gravitational collapse
to a stable pancake-like configuration as $ t$ goes from $0 \to\infty$. 

Thus far we have studied timelike geodesics in simple anisotropic cosmological models involving simultaneous collapse along one spatial direction and expansion along the other two spatial directions. It is interesting to note that some aspects of the geodesic phenomena under consideration here can be illustrated even in the standard spatially homogeneous and isotropic cosmological models. To this end, we consider, for simplicity, the Einstein-de Sitter model.

 \section{Einstein-de Sitter model}  
The metric of the Einstein-de Sitter universe is
\begin{align}
\label{eq:16} ds^2 =-dt^2+a^2(t)(dx^2+dy^2+dz^2),
\end{align}
where $a(t)=(t/t_0)^{2/3}$ and $t_0$ denotes the present epoch.
The source here is comoving dust ($P=0$) with density 
$ \rho=(6 \pi G t^2)^{-1}$.
The same procedure as above gives, for \emph{peculiar} velocity of a free test particle relative to the Hubble flow,
\begin{align}
\label{eq:18} 
\begin{split}
v_x=&\frac{C_1}{\sqrt{a^2+C^2}},\quad v_y=\frac{C_2}{\sqrt{a^2+C^2}},\\
v_z=&\frac{C_3}{\sqrt{a^2+C^2}},\quad \gamma=\sqrt{1+\frac{C^2}{a^2}}, 
\end{split}
\end{align}
where
$C^2=C_1^2+C_2^2+C_3^3$. As $t\to \infty$ ,  $a\to \infty$,  $\gamma\to 1$, $v_x\to 0$, $v_y\to 0$ and $v_z\to 0$.
Moreover, as $t\to 0$, $a \to 0$ and $\gamma\to \infty$; that is,  $v^2\to 1$ with
\begin{align}
\label{eq:19} 
v_x\to \frac{C_1}{|C|}, \quad  v_y\to \frac{C_2}{|C|}, \quad v_z\to \frac{C_3}{|C|} .
\end{align}

\section{discussion} 
The spacetimes considered in this Letter  involve similarity solutions of general relativity~\cite{ 12,13,14}.  For the gravitomagnetic jets,  for instance,  the generator of homothety is   $\xi=\partial_t+z \partial_z$,  so that 
we have $\xi_{\mu;\nu}+\xi_{\nu;\mu}=g_{\mu\nu}$.
Our examples are neither spatially compact nor asymptotically flat---these are general features of similarity solutions of general relativity according to~\cite{15}. The observational evidence in support of self-similarity in
astrophysical jets has been discussed in~\cite{16,17,18} and the references cited
therein. It remains to see if self-similarity will turn out to be a
characteristic feature of all dynamic gravitational fields that exhibit  ``acceleration"/``deceleration" phenomena.

According to the modern version of the Kant-Laplace nebular hypothesis, formation of structure in the universe generally follows from the collapse of a swirling cloud of gas and dust. The rotating and contracting cloud is characterized by a rotation axis along which matter collapses to a disk and thus there is contraction along the rotation axis and expansion along the other two spatial axes. In extreme situations, illustrated in this work via exact solutions of
general relativity, this circumstance can lead to the ``acceleration"  of free
particles to the speed of light relative to the ambient medium, resulting in
jets parallel and antiparallel to the rotation axis. We speculate that in
realistic situations the underlying physics should persist thus forming the
usual astrophysical double-jet configuration.  That is, when matter undergoes gravitational collapse, nearby free test particles moving in the
resulting dynamic gravitational field can accelerate up and down parallel to the
direction of collapse to form a double-jet structure with respect to the
collapsing system. This conception of the origin of cosmic jets has been discussed
here in the context of certain exact solutions of general
relativity, but may apply more generally.  In terms of the Newtonian gravitational potential, the contraction along the rotation axis produces negative Newtonian gravitational energy, which can in part be balanced by the positive kinetic energy of nearby free test particles moving up and down parallel to the rotation axis and accelerating to high speeds  with respect to the ambient collapsing medium. The more extreme the collapse is, the higher the expected jet speed, so that
in the ultimate limit of the collapse to an infinitely thin disk the jet
speed would then approach the speed of light in accordance with the exact
solutions of the gravitational field equations discussed in this work. It is important to emphasize that this purely gravitational mechanism for jet formation only operates while gravitational collapse is taking place. The
jets are up and down geodesic outflows normal to the plane of the resulting accretion disk.
  Most of these free particles escape as cosmic rays, but the ones near the core may encounter a strong magnetic field that could help to confine and  collimate them into narrow astrophysical jets along the rotation axis. To maintain such an initial double-jet configuration over time, however, would necessitate an appropriate MHD environment. These ideas may provide the physical basis for the origin of jets in
star-forming regions~\cite{ 20}.

\begin{acknowledgements}
BM is grateful to V. Belinski and H. Quevedo for helpful discussions. CC was supported in part by the NSF grant DMS 0604331.
 \end{acknowledgements}

\end{document}